\documentclass[%
 reprint,
 amsmath,amssymb,
 aps
]{revtex4-1}

\usepackage{graphicx}
\usepackage{dcolumn}
\usepackage{bm}
\usepackage{units}
\usepackage{amsmath}
\usepackage{mathtools}

\usepackage{color}

\begin{document}


\title{Cooling a Bose gas by three-body losses}
\author{Max Schemmer}
\author{Isabelle Bouchoule}%
 \email{isabelle.bouchoule@institutoptique.fr}
\affiliation{Laboratoire Charles Fabry, Institut d'Optique, CNRS, Universit\'e Paris Sud 11,
2 Avenue Augustin Fresnel, F-91127 Palaiseau Cedex, France}%


\begin{abstract}
  We report the first demonstration of cooling by three-body losses in a
Bose gas. We  use a harmonically confined one-dimensional (1D) Bose gas in the quasi-condensate
regime and,  as the atom number decreases under the effect of three-body losses,
the temperature $T$ drops up to
a factor four. The ratio $k_B T /(m c^2)$ stays close to 0.64, where $m$ is the 
atomic mass and $c$ the sound speed in the trap center.
The
dimensionless 1D interaction parameter $\gamma$, evaluated at the trap center, spans more than two
order of magnitudes over the different sets of data.  We present a theoretical
analysis for a homogeneous 1D gas in the quasi-condensate regime, which predicts that
the ratio $k_B T/(mc^2)$ converges towards 0.6 under the effect of three-body losses.
More sophisticated  theoretical predictions that take into account
the longitudinal harmonic confinement and transverse effects are in agreement within
30\% with experimental data.
\end{abstract}


\maketitle


The identification and understanding of cooling
processes, both on the theoretical and the experimental side, is
crucial to the development of cold atom
physics~\cite{dalibard_laser_1989,anderson_observation_1995}.
It can help to elaborate strategies to enter new regimes and it can also improve the control over
state preparation in experiments 
  where cold atoms are used as quantum simulators of many body systems.
Ultra-cold atom gases are metastable systems, their ground state
being a solid phase. 
They are thus plagued with intrinsic 
recombination processes, that in practice limit their lifetime.
Such process are mainly three-body collisions 
during which a strongly bound dimer is formed. It amounts to three-body
losses because the dimer is typically no longer trapped and the
remaining atom escapes because of its large kinetic energy.
These losses are known to produce an undesired  heating in cold gases.
In the case of a thermal gas, since they occur predominantly in the regions of high
atomic density, where the potential energy is low, these losses
increase the energy per remaining particle, leading to an 
anti-evaporation process~\cite{weber_three-body_2003}.
In Bose-Einstein condensates (BEC) confined in deep traps, it was predicted that three-body collisions produce a heating
of the BEC through secondary collisions with high energy excitations
formed by the loss process~\cite{guery-odelin_excitation-assisted_1999}.
This paper constitutes a breakthrough since for the first time we identify a cooling
associated to three-body losses in a cold Bose gas. 

The effect of losses has been investigated for  1D  Bose gases in the quasicondensate
regime~\footnote{Quasi-condensates are
  characteristic of weakly interacting 1D Bose gases at 
low enough temperature: repulsive interactions 
prevent large density fluctuations 
such that the gas resembles locally 
a BEC, although it does not sustain true long range order~\cite{petrov_regimes_2000}.} in the case of one-body
losses~\cite{rauer_cooling_2016,grisins_degenerate_2016,johnson_long-lived_2017,schemmer_monte_2017}.
This work was recently extended~\cite{bouchoule_cooling_2018-1} to 
  any j-body loss process, for Bose gases in the
BEC or quasicondensate regime,
in any dimension $d$, and for homogeneous gases as well as
gases confined in a smooth potential.
Theses  studies focus on the effect of losses 
on low energy excitations in the gas, the 
phononic modes, which  correspond to density waves propagating in 
  the condensate.
On the one hand, the energy in these modes
is reduced by losses since the amplitude of density modulations
is decreased, removing interaction energy from the mode. On the
other hand, the discrete nature of the loss process comes with accompanying
shot noise which induces density fluctuations, increasing
the energy per mode.
It has been shown that the competition
between these processes leads to a stationary value of the
ratio $k_B T / (m c^2)$ where $m$ is the atom mass and $c$ the
speed of sound.
This value, of the order of one, depends
on $j$, $d$,
and on the confining potential~\cite{bouchoule_cooling_2018-1}. For three-body losses in a 1D quasicondensates
($j=3$, $d=1$)  confined in a harmonic potential one expects $k_B T/(mc_p^2)$ to converge to
0.70~\cite{bouchoule_cooling_2018-1}, where $c_p$ is evaluated at the peak density.

In this paper, we show  
for the first time experimentally that three-body losses induce a cooling and we identify the
stationary value of  $k_B T/(mc_p^2)$ associated to the three-body process.
More precisely, investigating the time evolution of a 1D quasi-condensate, we  
observe a decrease of  the temperature  as the atom number decreases under the
effect of three-body losses. 
Moreover, on the whole observed time-interval, the ratio 
 $k_BT/(mc_p^2)$ stays about constant, at a value close to $0.64$,
which indicates that the  stationary value of $k_B T/(mc_p^2)$ imposed by the loss process
is reached. 
We took several data sets for different parameters. 
In terms of the 1D dimensionless parameter $\gamma$~\cite{lieb_exact_1963}
characterizing the strength 
  of the interactions~\footnote{$\gamma=mg/(\hbar^2n)$ 
  where $g$ is the interaction
  coupling constant, $n$ the linear density, $m$ the atomic mass and $\hbar$ the Planck constant. We
  evaluate it using the linear density at the trap center.},
 our data
span more than two orders of magnitude. 
We compare the experimental data with numerical calculations based on the
results of~\cite{bouchoule_cooling_2018-1},
which take into account
the harmonic longitudinal confinement of the gas and the swelling of the transverse wave function under
the effect of interactions.
The experimental results are close to those predictions. 
In order to present the underlying physics, we derive in this paper
the evolution of the temperature
under three-body losses, in the more simple case of a homogeneous purely 1D quasicondensate.

The experiment uses an atom-chip set
up~\footnote{The experiment is described in more detail in~\cite{jacqmin_momentum_2012}.} 
where  $^{87}$Rb atoms are magnetically confined using
current-carrying micro-wires.
An elongated atomic cloud is prepared using radio
frequency (RF) forced evaporative cooling
in a trap  of transverse frequency
$\omega_{\perp}$. 
  Depending on the data set,  $\omega_{\perp}/(2\pi)$ varies between \unit[1.5]{kHz} and  \unit[9.2]{kHz} and
  the atomic peak
linear densities $n$ vary between $\unit[22]{\mu m^{-1}}$ and $\unit[257]{\mu m^{-1}}$.
The temperature fulfills $k_BT<\hbar\omega_\perp$ and  
the gas mostly
behaves as a 1D Bose gas~\cite{armijo_mapping_2011}. It moreover   
lies in the quasicondensate regime~\cite{kheruntsyan_pair_2003}, characterized 
by weak correlations between atoms, as in a Bose-Einstein condensates~\footnote{
Thermally activated  phonons  however
prevent the establishment of a well defined phase.}, and in particular
small density fluctuations 
~\footnote{Density fluctuations are
considered only in a  coarse grained approximation, valid 
for lengths much larger than the
interparticle distance.}.
As long as the atoms are in the ground state of the transverse potential,
interactions between atoms are well described by a 1D effective coupling 
constant
$g=2\hbar\omega_\perp a$, where $a=$\unit[5.3]{nm} is the 3D scattering length~\footnote{
We are far from confinement-induced resonances predicted in~\cite{olshanii_atomic_1998}.}, 
and the chemical potential
is given by $\mu=gn$. 
This is valid only as
long as $\mu\ll \hbar \omega_\perp$, which requires $na\ll 1$.
In the presented data $na$ takes values as large as 1.3 and the broadening of the transverse
wavefunction due to interactions has to be taken into account for quantitative analysis.
In particular, the equation of state becomes
  $\mu = \hbar \omega_{\perp} (\sqrt{1+4na} - 1)$~\cite{fuchs_hydrodynamic_2003}.
The quasi-condensates is 
confined in the longitudinal direction with 
a harmonic potential $V(z)$ of trapping
frequency $\omega_z/(2\pi)=\unit[8.5]{Hz}$, weak enough so that
the longitudinal profile $n_0(z)$ is well described by the 
Local Density Approximation (LDA), with
a local chemical potential $\mu(z)=\mu_p-V(z)$, where $\mu_p$ is the
peak chemical potential.  It extends over $2R$ 
where the Thomas-Fermi radius $R$ fulfills $V(R)=\mu_p$.
Once the quasicondensate is prepared, we increase the frequency of the 
radio-frequency field, by several kHz, a value sufficient
so that it no longer induces losses. We then investigate the evolution
during the waiting time $t$. 
Five different data set are investigated, differing in the value of the 
transverse confinement and the initial temperature and peak density.

\begin{figure}[h!]
\centerline{\includegraphics[width=\linewidth]{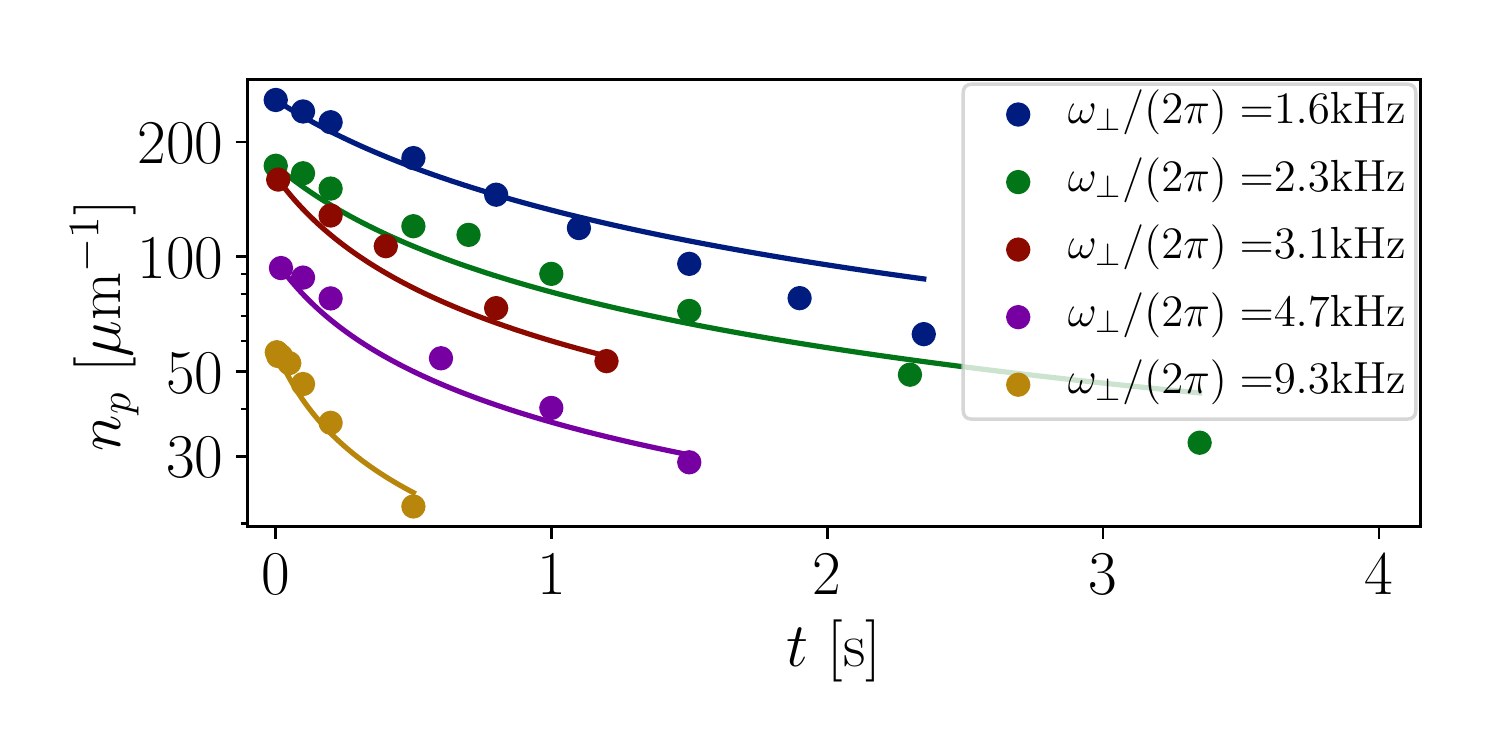}}
\caption{Peak  density, in log scale, versus the waiting time $t$,
    for the five different data-sets. Solid lines are
    ab-initio calculations of the effect of three-body losses, for initial peak densities equal to
    that of the experimental data.}
\label{fig.n_p}
\end{figure}

Using absorption images we record the density profile of the gas, from
which we extract the peak density $n_p$.
Fig.~\ref{fig.n_p} shows evolution of 
  $n_p$ with the waiting time $t$
for the different data sets.  
We observe a decrease of $n_p$ whose origin is three-body recombinations,
as justified by calculations
presented below.
In a three-body recombination, a molecule (a dimer) is formed and its binding
energy is released in the form of kinetic energy of the molecule and
the remaining atom. They both
  leave the
trap since their energy is typically much larger than the trap depth,
limited by 
  the radio-frequency field.
Thus, the effect of three-body process is to decrease the 
gas density according to  ${\rm d}\rho/{\rm d}t= - \rho^3g^{(3)}(0)\kappa$, 
where $\rho$ is the three dimensional atomic density,  
$g^{(3)}(0)$ is the normalized three-body correlation function at zero
distance, and $\kappa = \unit[(1.8\pm 0.5)\times 10^{-41}]{m^6/s}$
is the three-body loss rate for $^{87}$Rb~\cite{soding_three-body_1999}. 
In a quasi-condensate, correlations between atoms are 
small and $g^{(3)}(0)\simeq 1$~\footnote{According to 1D
  Bogoliubov calculations, $g^{(3)}(0)-1=3(g^{(2)}(0)-1)=\sqrt{\gamma}f(k_BT/(gn))$,
  where $f$ is
  a dimensionaless function wich takes the value $f(0.64)\simeq -1.5$.
  We then find {\it a posteriori}
  that $|g^{(3)}(0)-1|$ spans the interval $[0.02,0.25]$ for our data.}. 
Moreover, integrating ${\rm d}\rho/{\rm d}t$ over the transverse shape of 
the cloud, we 
obtain a  one-dimensional rate of density decrease
$  {\rm d}n_0(t)/{\rm d}t = - K n_0(t)^3$, 
where $K  = \kappa/n_0^3 \iint {\rm d} x {\rm d} y \, \rho(x,y)^3 $.
Taking into account the transverse broadening of the wavefunction using the Gaussian
ansatz results of \cite{salasnich_effective_2002}, we obtain 
$K=K^0/(1+2n_0a)$, where 
$K^0=\kappa  m^2 \omega_{\perp}^2/(3 \pi^2 \hbar^2 ) $
\footnote{We check the Gaussian ansatz 
gives correct results
up to 20\% for our parameters by comparing  with numerical
solution of the Gross-Pitaevskii equation.}.
Finally,
the rate of variation of 
the total atom number $N$ is
\begin{equation}
{\rm d} N/{\rm d}t = -\int_{-R}^{R}{\rm d}z K(z) n_0(z)^3.
\label{eq.dNdt}
\end{equation}
At any time, the measured profile is very close to an equilibrium profile,
  which indicates the loss rate
  is small enough to ensure adiabatic following of $n_0(z)$. Then
  $N$ and $n_0(z)$ are completely determined by 
  $n_p$ and
Eq.~(\ref{eq.dNdt}) can be transformed into a 
differential equation for $n_p$.
We solve it numerically
for the parameters of the experimental data, 
namely the frequency $\omega_\perp$ and the initial peak density,
using the LDA to rely $N$ and $n_0(z)$ to $n_p$.
Calculations, shown in Fig.~\ref{fig.n_p}, are in good agreement 
with the experimental data, which confirms that losses are largely dominated by three-body losses.

\begin{figure}[h!]
\centerline{\includegraphics[width=\linewidth]{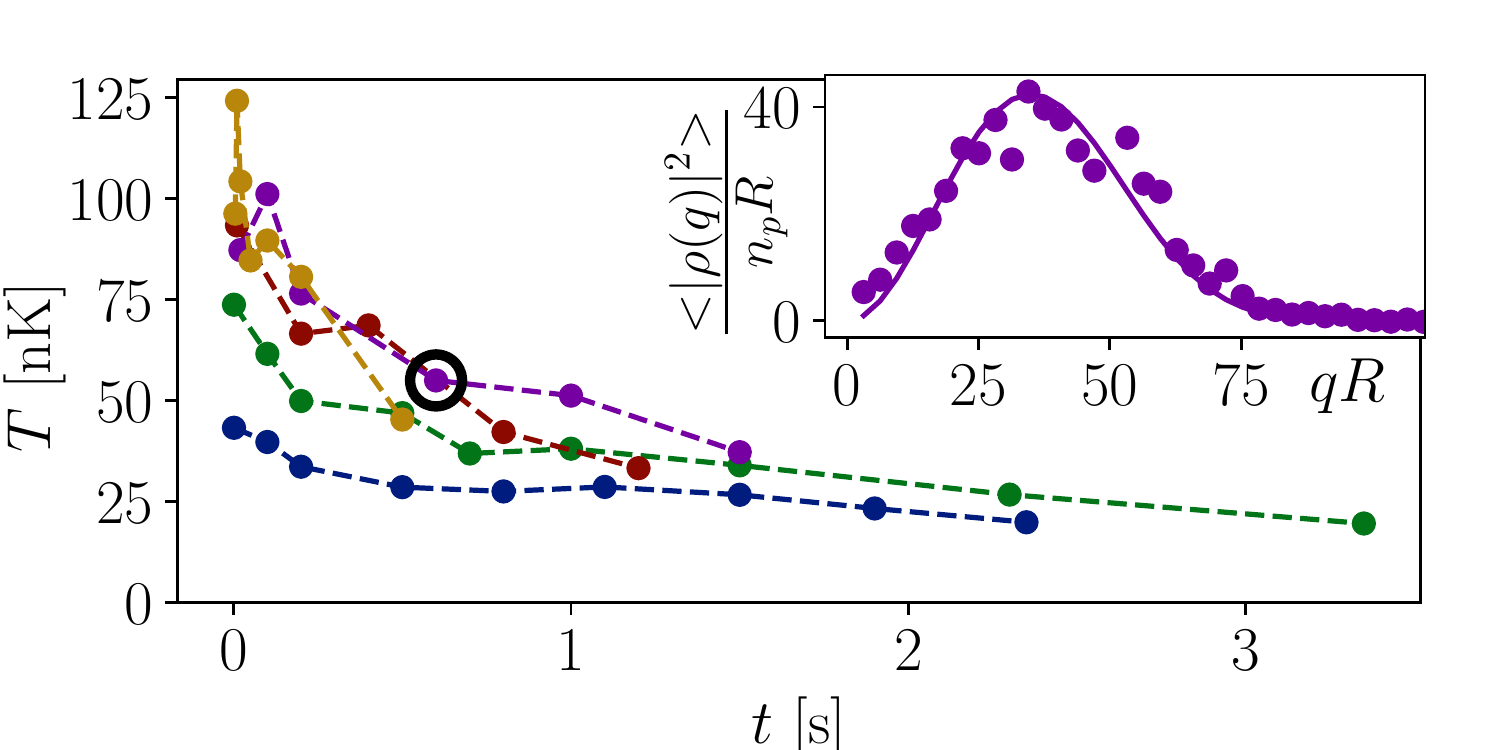}}
\caption{Evolution of the temperature for the five data sets (same color code as in Fig.~\ref{fig.n_p}).
    Inset:  density ripples power spectrum corresponding to the encircled point, with the fit in solid line yielding the temperature. }
\label{fig.T_T0}
\end{figure}

The temperature of the gas is determined analyzing the large density ripples that appear after a 
time of flight $t_f$~\cite{imambekov_density_2009,dettmer_observation_2001,manz_two-point_2010,rauer_cooling_2016,schemmer_monitoring_2017}. Interactions are effectively quickly turned off by the transverse expansion of the gas 
and the subsequent free evolution transforms   
longitudinal phase fluctuations 
into density 
fluctuations.
Using an ensemble of images taken in the same experimental condition, we
extract the density ripple power spectrum  
\begin{equation}
\langle |\rho(q)|^2\rangle =  
\left \langle 
\left | \int  {\rm d} z \left(n(z,t_{f})-\langle n(z,t_{f})\rangle \right) e^{iqz} \right |^2
\right \rangle. 
\end{equation}
We choose
$t_f$  small enough so that the 
density ripples occurring near the position $z$ are
produced by atoms which where initially 
in a small portion of the cloud, located near $z$.
We can thus use, within a LDA, 
the analytic predictions for homogeneous gases
to compute the expected power spectrum of the trapped gas~\cite{schemmer_monitoring_2017}.
We take into account the finite resolution of the imaging system
modeling its impulse response function by a Gaussian of rms wifth $\sigma_{\rm{res}}$.
For a given data set the density ripple power spectrum recorded at $t=0$ is fitted with the temperature $T$ and $\sigma_{\rm{res}}$, the latter depending on the 
transverse width of the cloud and thus on $\omega_\perp$.
We then fit $\langle |\rho_q|^2\rangle$ at larger values of $t$
with  $T$ as a single parameter (see inset Fig.~\ref{fig.T_T0}). 
The time evolution of $T$  is shown in 
Fig.~\ref{fig.T_T0} for the five different data sets investigated in this paper.
The  temperature decreases with $t$, which 
indicates a cooling mechanism associated to the three-body losses.
Note that this thermometry probes  phononic
 collective modes since 
 the experimentally accessible wavevectors are much smaller
 than the inverse healing length $\xi^{-1} = \sqrt{m g n_0}/\hbar$.

\begin{figure}[htbp]
\centerline{\includegraphics[width=\linewidth]{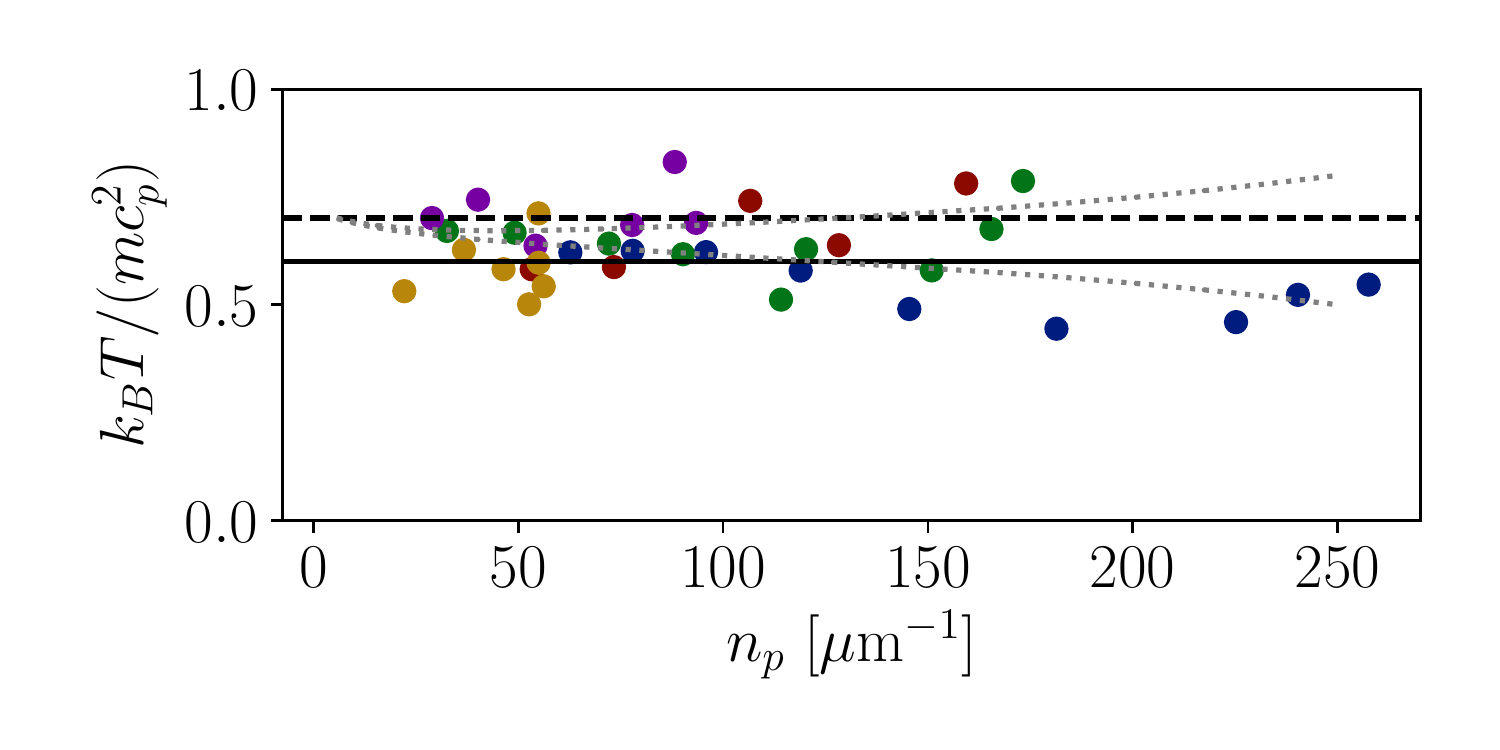}}
\caption{Evolution of 
    the ratio $k_B T /(mc_p^2)$, in the course of the three-body loss process, for the five data sets (same color code
    as in Fig.~\ref{fig.n_p}).
    Solid (resp. dashed) lines: asymptotic ratio for a 1D  homogeneous (resp.  harmonically confined) gas.
    Dotted lines: numerical calculation, that take into account
    the transverse swelling, for two different initial situations
    closed to that of experimental data. 
    }
\label{fig.k_BT_mc2}
\end{figure}

Fig.~\ref{fig.k_BT_mc2} shows the same data, with the temperature normalized 
to $mc_p^2$, where $c_p=\sqrt{n_p \partial_n\mu|_{n_p}/m}$ is the sound velocity at the center of the cloud,
shown versus the peak density $n_p$.
While $n_p$  explore more than one order of magnitude, remarkably $k_BT/(mc_p^2)$
shows small dispersion and 
is close to its mean value $0.64$, the standard deviation being $0.02$~\footnote{These values
    have been obtained on the data-points satisfying $n_p a < 0.2$ such that
    the effect of transverse swelling is small.}.

\begin{figure}[htbp]
\centerline{\includegraphics[width=\linewidth]{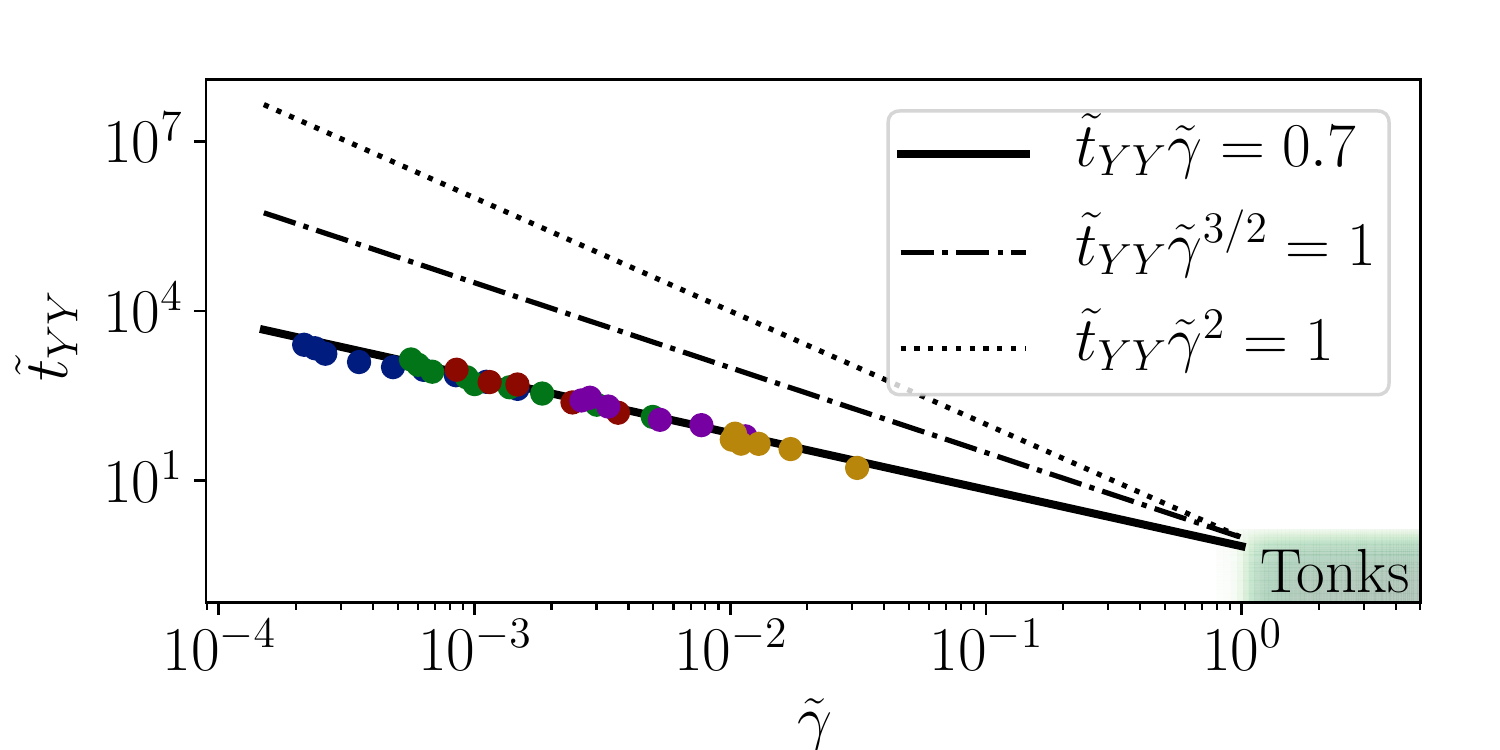}}
\caption{The data collapse on the line $\tilde{t}_{YY} \tilde \gamma = 0.7$.
  The lower right corner corresponds to the strongly interacting Tonks-Girardeau regime.
  The data sets and color codes are the same as in all other figures.}
\label{fig.phase_diagram}
\end{figure}

The absolute linear density is however not the most relevant
quantity. 
A 1D gas at thermal equilibrium is characterised by the dimensionless
quantities $\gamma=mg/(\hbar^2 n)$ and
$t_{\rm{YY}}= \hbar^2 k_B T/(m g^2)$~\cite{kheruntsyan_pair_2003}. 
In particular the quantum degeneracy condition corresponds to the line $t_{\rm{YY}} \gamma \simeq 1$.
Moreover, the
crossover between the ideal Bose gas regime 
and the quasicondensate regime
occurs, within the region $\gamma \ll 1$, along the line 
$t\gamma^{3/2}\simeq 1$.
Finally, within the quasi-condensate regime, the
  line $t_{\rm{YY}}\gamma \simeq 1$ separates the high temperature regime,
  where the zero distance two-body correlation function $g^{(2)}(0)$ 
  is dominated by thermal fluctuations and $g^{(2)}(0)>1$,
  from the low temperature regime, where $g^{(2)}(0)$ is dominated by quantum fluctuations
  and is smaller than 1~\footnote{$g^{(2)}(0)=1$ for $t_{\rm{YY}}\gamma = 1.5(1)$.}.
 Here we generalize these 1D parameters to quasi-1D gases introducing 
 $\tilde{t}= \hbar^2 k_B T n^2/(m^3c^4)$ and $\tilde{\gamma}=m^2c^2/(\hbar^2n^2)$.
For a harmonically confined gas, we refer in the following to
    the values of $\tilde{t}$
    and $\tilde{\gamma}$ evaluated at the trap center.
The evolution of the state of the gas during the three-body loss process is shown 
in Fig.~\ref{fig.phase_diagram} in the ($\tilde{t},\tilde{\gamma}$) space. All 
data collapse on the line $\tilde{t}\tilde{\gamma}= k_B T/(m c^2_p)= 0.7$, with a maximum deviation of
36\%, while $\tilde{t}$ explore more than 2 order of magnitude.

The physics at the origin of the observed behavior can be understood by
considering the simple case of a pure 1D homogeneous quasicondensate.
We give here a simplified analysis and refer the reader to~\cite{bouchoule_cooling_2018-1} for a more
complete study.
At first,
let us solely consider the effect of three-body losses, during a time interval $dt$,
in a small cell of the gas of length $\Delta$. 
The  density is $n=n_0+\delta n$, where $n_0$ is the mean density and
$\delta n\ll n_0$ since we consider a quasicondensate.
The  density evolves
according to $d n = - K n^3 dt + d\eta $, where $d \eta$ is
a random  variable of vanishing mean value reflecting the stochastic nature of the loss process.
During $dt$ the loss process is close to poissonian
and $\langle d \eta^2\rangle = 3 K n^3/\Delta d t\simeq 3 K n_0^3/\Delta dt$, where
the factor 3 comes from the fact that each loss event amounts to the loss of 3 atoms. 
To first order in $\delta n$,
  the mean density evolves according to $d n_0 = - K n_0^3 dt $,  and
expansion of $dn$  yields
\begin{equation}\label{eq.deltan}
  {\rm d} \delta n = - 3 K n_0^2 \delta n {\rm  d}t + {\rm d} \eta.
\end{equation}
The two terms of the r.h.s. correspond to the two competing effects of losses.
The first term, a drift term, reduces the density fluctuations: it thus decreases the
interaction energy, leading to a cooling.
The second term, a stocastic term due to  the discrete nature of the atom losses,
increases the density fluctuations and thus induces a heating.
Going to the continuous limit, one has $\langle d\eta(z)d\eta(z')\rangle=3K n_0^3 dt\delta(z-z')$.
 
Let us now consider the intrinsic dynamics of the gas.
Within the Bogoliubov approximation, valid in the quasicondensate regime,
one identifies independent collective modes and, up to a constant term,
the Hamiltonian of the gas writes $H = \sum_k H_k$, where
\begin{equation}
H_k = A_k \delta n_k^2 + B_k \theta_k^2
\end{equation}
is the Hamiltonian of the collective mode of wave vector $k$~\cite{schemmer_monte_2017}.
Here the conjugate quadratures
$\delta n_k$ and $\theta_k$ are the Fourier components of $\delta n$ and $\theta$,
$B_k = \hbar^2 k^2 n_0/(2m)$ and, as long as phononic modes are considered, $A_k = g/2$.
At thermal
equilibrium the energy is equally distributed between the quadratures so that
$\langle H_k\rangle/2 = A_k \langle \delta n_k^2 \rangle = B_k
\langle\theta_k^2\rangle $.
Let us compute the evolution of $\langle H_k\rangle$ under the effect of losses,
assuming the loss rate is small compared to the
mode frequency $\omega_k$ such that the equipartition
holds for all times. 
First the Hamiltonian parameter $B_k$ changes according to 
$d B_k = -K n_0^2 B_k d t$.
Second, according to Eq.~\ref{eq.deltan}, the losses modifies the distribution 
on the quadrature $\delta n_k$ and we obtain
$d\langle \delta n_k^2\rangle/dt= -6Kn_0^2 \langle\delta n_k^2\rangle + 3K n_0^3$~\footnote{Losses
  are also responsible for an increase of the 
  spread along the quadrature $\theta_k$. However, the associated heating is negligible 
  for phononic modes~\cite{bouchoule_cooling_2018-1}.}.
 Summing this two contributions leads to 
\begin{equation} 
\frac{d \langle H_k\rangle}{ dt} = - 7/2 Kn_0^2 \langle H_k\rangle  + 3/2 K n_0^3 g.
\end{equation}
From this equation, and using $dn_0/dt=-Kn_0^3$, we derive the evolution of the ratio
$y=\langle H_k\rangle/(mc^2)$, where $c=\sqrt{gn_0/m}$ is the speed of sound.
We find that $y$ converges at long times towards the
stationary value  $y_\infty=0.6$.
Phononic modes
  typically have large occupation numbers for values of $y$ of the order of or larger than 1 so that
$\langle H_k \rangle \simeq k_B T$, where $T$ is the mode temperature,
and  $y  = k_B T / (m c^2)$.

In presence of a harmonic longitudinal potential, calculations
  which assume that the loss rate is small enough to neglect non-adiabatic coupling between modes,
  predict a stationary value of  the ratio $k_B T/(mc_p^2)$ equal to
  $y_\infty=0.70(1)$~\cite{bouchoule_cooling_2018-1}, a value close to 
  experimental data.
  For a more precise comparison of data with theory, we compute the time-evolution of $y$
  according to formula derived in~\cite{bouchoule_cooling_2018-1}, that take into account
  the transverse swelling of the wavefunction which occurs in our data at large $na$.
The results, shown in Fig.~\ref{fig.k_BT_mc2} for two different initial situations, is close to
experimental data.
Even at the beginning of the  observed time-evolution, the ratio $k_BT/(mc_p^2)$ in  our gases
  is close to its asymptotic value.
Data are taken only for gases that were  sufficiently cooled by evaporative cooling to be in the
quasicondensate regime, where both our thermometry  and the theoretical description
of the effect of losses are applicable. It occurs that, in our experiment, when the gas enters
the quasicondensate regime the ratio $k_BT/(mc_p^2)$ is already close to 0.7.

In conclusion, we showed in this paper 
that, under a three-body losses process, the temperature
of a quasicondensate in the quasi-1D regime  decreases in time. The
ratio $k_BT/(mc_p^2)$ stays close to 
the predicted  stationary
value, which results from the competition between the cooling effect of losses
and the heating due to the stochastic nature of losses. 
This work raises many different questions.
First the cooling  mechanism presented in this paper is not restricted to 1D quasicondensates and it
would be interesting to investigate it in other regimes and dimensions,  in particular
as one approaches the Tonks regime of 1D gases. 
Second, while results presented in this paper 
  concern only the phononic modes, it would be interesting to
  study the effect of losses on  higher energy modes. They might
  reach  higher 
  temperatures than phononic modes, as predicted for  one-body 
  losses~\cite{johnson_long-lived_2017}, and the stability of such a non thermal situation might be
  particular to the case of 1D gases. 
Finally, it is interesting to compare the  three-body losses cooling to
the commonly used evaporative cooling mechanism, which 
occurs via the removal of atoms 
whose energy energy is larger than the trap depth.
Its efficiency drops drastically for temperatures lower than $mc_p^2/k_B$: the
relevant excitations are then phonons, which  do not extend
beyond the condensate, 
and are thus very difficult to ``evaporate''. 
Thus, obtaining, by means of evaporative cooling,
temperatures lower than the asymptotic temperature imposed by three-body losses
is not guaranteed.

This work was supported by R\'egion
\^{I}le de France (DIM NanoK, Atocirc project). 
The authors thank Dr Sophie Bouchoule of C2N (centre nanosciences 
et nanotechnologies, CNRS / UPSUD, Marcoussis, France) for the development 
and microfabrication of the atom chip. Alan Durnez and Abdelmounaim Harouri 
of C2N are acknowledged for their technical support. C2N laboratory is a 
member of RENATECH, the French national network of large facilities for 
micronanotechnology.
M. Schemmer acknowledges support by the Studienstiftung des Deutschen Volkes.

%

\end{document}